\documentclass[aip,jcp,reprint]{revtex4-1}

\usepackage{graphicx}
\usepackage{amsmath}
\usepackage{amssymb}
\usepackage{multirow}

\newcommand{\Angstrom}{\ensuremath{\mathring{\textnormal{A}}}}
\newcommand{\sub}[1]{\ensuremath{_{\textrm{#1}}}} 
\newcommand{\super}[1]{\ensuremath{^{\textrm{#1}}}} 
\newcommand{\sci}[2]{\ensuremath{#1 \times 10^{#2}}}

\begin{document}

\title{The charge-asymmetric nonlocally-determined local-electric (CANDLE) solvation model}

\author{Ravishankar Sundararaman and William A. Goddard III}
\affiliation{Joint Center for Artificial Photosynthesis, California, USA}
\date{\today}

\begin{abstract}
Many important applications of electronic structure methods involve molecules
or solid surfaces in a solvent medium. Since explicit treatment of the solvent
in such methods is usually not practical, calculations often employ
continuum solvation models to approximate the effect of the solvent.
Previous solvation models either involve a parametrization based on atomic radii,
which limits the class of applicable solutes, or based on solute electron density,
which is more general but less accurate, especially for charged systems.
We develop an accurate and general solvation model that includes a cavity
that is a nonlocal functional of both solute electron density and potential,
local dielectric response on this nonlocally-determined cavity,
and nonlocal approximations to the cavity-formation and dispersion energies.
The dependence of the cavity on the solute potential enables
an explicit treatment of the solvent charge asymmetry.
With only three parameters per solvent, this `CANDLE' model simultaneously reproduces
solvation energies of large datasets of neutral molecules, cations and anions
with a mean absolute error of 1.8~kcal/mol in water and 3.0~kcal/mol in acetonitrile.
\end{abstract}

\maketitle

\section{Introduction} \label{sec:Introduction}

Solvents play a critical role in determining chemical reaction mechanisms and rates,
but the need for thermodynamic phase-space sampling renders direct treatment
of the liquid in electronic structure calculations far too computationally intensive.
The standard solution to this problem is to use continuum solvation
models which empirically describe the dominant effects
of the solvent within a single electronic structure calculation of the solute alone.
This enables rapid estimations of free energies of reaction intermediates,
allowing for a theoretical screening of reaction mechanisms,
and providing insight into the mechanisms involved in catalysis
required for the development of more efficient catalysts.

Conventional continuum solvation models such as the `SM' series \cite{PCM-SM1,PCM-SM8,PCM-SMD}
and the polarizable continuum models (PCMs) \cite{PCM94,PCM97,PCM-Review}
construct cavities composed of a union of van der Waals (vdW) spheres centered on the solute atoms,
and approximate the effect of the solvent by the electric response
of a continuum dielectric cavity along with empirical corrections
for cavity formation and dispersion energies.
These models include a number of atom-dependent parameters such as radii
and effective atomic surface tensions which are fit to datasets of
experimental solvation energies, typically including neutral and charged organic solutes.
These models can be quite accurate for the solvation energies of solutes similar
to those in the fit set, but require care when extrapolating to new systems.
Additionally, the sharp cavities generated from the union of atomic spheres
can lead to numerical difficulties including non-analyticities in the energy landscape
for ionic motion that complicates geometry optimization and molecular dynamics of the solute.

In contrast, density-based solvation models such as the self-consistent continuum solvation
(SCCS) approach \cite{PCM-SCCS,PCM-SCCS-charged} and the simplified solvation models 
\cite{PCM-Kendra,NonlinearPCM,CavityWDA} within joint-density functional theory (JDFT) \cite{JDFT}
employ a continuously varying dielectric constant determined from the solute electron density.
These models avoid the numerical difficulties arising from sharp spheres making them
more naturally suited for the plane-wave basis sets used in solid-state calculations.
Additionally, density-based solvation models typically involve fewer (two to four) parameters
and should extrapolate more reliably from one class of solute systems to another.
However, the smaller parameter set also limits
the typical accuracy achievable in this class of models.
In particular, these solvation models exhibit a systematic error
between the solvation of cations and anions, with cations in water
over-solvated and anions under-solvated.
This issue is sometimes handled by fitting separate
parameter sets for differently-charged solutes,\cite{PCM-SCCS-charged}
but that is not an option for solutes that combine centers
of opposite charges such as zwitterions or ionic surfaces.

Here we report a highly accurate density-based solvation model
that addresses the aforementioned charge asymmetry issue. We start with
the recent non-empirical solvation model, `SaLSA',\cite{SaLSA} derived from
the linear-response limit of joint density-functional theory,\cite{JDFT}
which provides an excellent starting point due to the independence of its cavity
from fitting to solvation energies, and provides additional numerical stability
from the nonlocality in the determination of the cavity, the electric response,
cavity formation free energy and dispersion energy.
To account for the charge asymmetry, section~\ref{sec:CavityDetermination}
introduces a nonlocal dependence of the cavity on both the
electron density and electric potential of the solute.

In SaLSA, the nonlocal electric response involves an angular momentum expansion
which converges rapidly only for small sphere-like solvent molecules (such as water)
and which increases the computational expense by one-two orders of magnitude compared to local response.
Section~\ref{sec:ElectricResponse} replaces the nonlocal electric response with a local dielectric
as in traditional continuum solvation models, but derived from the nonlocal
cavity that builds in the charge asymmetry. Because it is based on the
Charge-Asymmetric Nonlocally-Determined Local Electric response,
we refer to this new model as the CANDLE solvation model.
The treatment of the cavity formation and dispersion energies are
almost identical to SaLSA,\cite{CavityWDA,SaLSA} except for minor
modifications to the dispersion functional (section~\ref{sec:Dispersion})
to improve the generalization to solvents of highly non-spherical molecules.
Finally, section~\ref{sec:Results} details the fits of the
three parameters in the model -- a charge asymmetry parameter,
an electric response nonlocality parameter and the dispersion scale factor
-- to experimental solvation energies. That section then demonstrates the accuracy of the
CANDLE solvation model for water and acetonitrile as prototypical solvents.

\section{Description of model} \label{sec:Model}

Following the SaLSA solvation model,\cite{SaLSA} we approximate the
total free energy of a solvated electronic system as
\begin{equation}
A\sub{sol}[n] = A\sub{HK}[n] + U\sub{lq}[\rho\sub{el},s] + G\sub{cav}[s] + E\sub{disp}[s].
\label{eqn:Asol}
\end{equation}
Here, $A\sub{HK}[n]$ is the Hohenberg-Kohn functional\cite{HK-DFT}
of the solute electron density $n(\vec{r})$, which in practice
we treat using the Kohn-Sham formalism\cite{KS-DFT} with an
approximate exchange-correlation functional. The second term
$U\sub{lq}[\rho\sub{el},s]$ is the electrostatic interaction energy
between the solute and solvent, where $\rho\sub{el}(\vec{r})$
is the total (electronic + nuclear) solute charge density
and $s(\vec{r})$ is the cavity shape function which switches
smoothly from 0 in the region of space occupied by the solute
to 1 in that occupied by the solvent. The third and fourth terms
of (\ref{eqn:Asol}) capture the cavity formation free energy
and dispersion energy respectively.

The following sections describe each of the above terms in detail.
Section~\ref{sec:CavityDetermination} presents the determination
of the cavity shape function $s(\vec{r})$, section~\ref{sec:ElectricResponse}
describes the electric response of the solvent that determines $U\sub{lq}$
and section~\ref{sec:Dispersion} details the dispersion energy $E\sub{disp}$.

For the cavity formation free energy $G\sub{cav}$, we adopt the parameter-free
weighted density approximation from Ref.~\citenum{CavityWDA} without modification.
Briefly, this model for the cavity formation free energy begins with a
weighted-density \emph{ansatz} motivated from an intuitive microscopic picture
of surface tension and completely constrains the functional form to bulk properties
of the solvent including the number density, surface tension and vapor pressure.
The resulting functional accurately describes the free energy of forming
microscopic cavities of arbitrary shape and size in comparison to
classical density-functional theory and molecular dynamics results.\cite{PolarizableCDFT}
(See Ref.~\citenum{CavityWDA} for a full specification of $G\sub{cav}[s]$.)

\subsection{Cavity determination} \label{sec:CavityDetermination}

Traditional density-based solvation models determine the cavity
as a local function of the solute electron density, $s(\vec{r}) = s(n(\vec{r}))$,
that switches from 0 to 1 over some density range (controlled by $n_c$ in the
JDFT simplified solvation models\cite{PCM-Kendra,NonlinearPCM} and by $(\rho\sub{min},\rho\sub{max})$
in the SCCS models\cite{PCM-SCCS,PCM-SCCS-charged}) that is fit to solvation energies.
In contrast, the non-empirical SaLSA model determines the cavity from an
overlap of the solute and solvent electron densities,
\begin{equation}
s(\vec{r}) = \frac{1}{2}\operatorname{erfc} \ln \frac{n\sub{lq}^0(r) \ast n(\vec{r})}{\bar{n}_c} \label{eqn:shapeSaLSA}
\end{equation}
where $n\sub{lq}^0(r)$ is a spherical average of the electron density of a single solvent molecule.
The critical density product $\bar{n}_c = \sci{1.42}{-3}$ is a universal solvent-independent constant
determined from a correlation between convolutions of spherical electron densities
of pairs of atoms and their van-der-Waals (vdW) radii. (See Ref.~\citenum{SaLSA} for details.)

We make two modifications to the SaLSA cavity determination.
First, the spherically-averaged electron density of the 
solvent molecule produces the correct cavity sizes for
the small approximately-spherical solvent molecules
(such as water, chloroform and carbon tetrachloride)
for which SaLSA works well. To generalize the approach beyond
such solvents, and to simplify the construction of the model so as
to not depend on electronic structure calculations of the solvent,
we replace the solvent electron density with a simple Gaussian model,
\begin{flalign}
n\sub{lq}^0(r) &\equiv Z\sub{val} w\sub{lq}(r)\nonumber\\
\textrm{with } w\sub{lq}(r) &\equiv \frac{1}{(\sigma\sub{lq}\sqrt{2\pi})^3}
	\exp\left(\frac{-r^2}{2\sigma\sub{lq}^2}\right).
\label{eqn:SolventGaussianModel}
\end{flalign}
Here, $Z\sub{val}$ is the number of valence electrons in the solvent molecule
and the Gaussian width $\sigma\sub{lq}$ is selected so that the overlap
of the model electron densities of two solvent molecules crosses $\bar{n}_c$
at a separation equal to twice the vdW radius $R\sub{vdW}$ of the solvent.
This condition reduces to the transcendental equation in $\sigma\sub{lq}$,
\begin{equation}
(n\sub{lq}^0 \ast n\sub{lq}^0)(2R\sub{vdW})
= \frac{Z\sub{val}^2}{(2\sigma\sub{lq}\sqrt{\pi})^3}
	\exp\left(\frac{-R\sub{vdW}^2}{\sigma\sub{lq}^2}\right)
= \bar{n}_c.
\end{equation}
Consistency of the above condition with the correlation between atom density overlaps
and atomic vdW radii,\cite{SaLSA} results in cavities of the appropriate size
(corresponding approximately to atomic spheres of radius equal to sum of
solute atom and solvent vdW radii).

Second, we modify (\ref{eqn:shapeSaLSA}) to account for the charge asymmetry in solvation.
Dupont and coworkers\cite{PCM-SCCS-charged} show that their characteristic solute electron
density parameters that fit the solvation energies of anions in water is an order of
magnitude larger than those that fit solvation energies of cations.
Thus they recommend separate parameter sets depending on the charge of the solute.
We consider this far too restrictive as it precludes applications
to solutes that combine sites with different charges.
Therefore we build in a dependence of the cavity on the solute electron potential that
effectively adjusts the critical electron density depending on the `neighborhood',
\begin{multline}
s(\vec{r}) = \frac{1}{2}\operatorname{erfc} \left[
	\ln \frac{Z\sub{val}\bar{n}(\vec{r})}{\bar{n}_c}
\right. \\ \left.
	- \operatorname{sign}(p\sub{cav})
		f\sub{sat}\left(\left|p\sub{cav}\right| \hat{e}_{\nabla\bar{n}}\cdot\nabla\hat{K}\bar{\rho}\sub{el}(\vec{r}) \right)
\right].
\label{eqn:shapeCANDLE}
\end{multline}
Here, $\bar{n} \equiv w\sub{lq} \ast n$ and
$\bar{\rho}\sub{el} \equiv w\sub{lq} \ast \rho\sub{el}$
are weighted electron densities and total charge densities respectively.
The remainder of this section specifies the remaining attributes of (\ref{eqn:shapeCANDLE}).

The combination $\nabla\hat{K}\bar{\rho}\sub{el}$ is the negative of
the electric field due to the solute (spatially-averaged by the
convolution with $w\sub{lq}$) since $\hat{K}$ is the Coulomb operator,
and $\hat{e}_{\nabla\bar{n}}$, the unit vector along $\nabla\bar{n}$,
is parallel to the inward normal of the cavity. Therefore,
the argument of $f\sub{sat}$ in (\ref{eqn:shapeCANDLE}) is proportional
to the spatially-averaged outward electric field due to solute,
which is negative for cation-like regions and positive for anion-like regions
(using an electron-is-positive sign convention for electrostatics).

Now note that we can write (\ref{eqn:shapeCANDLE}) as (\ref{eqn:shapeSaLSA})
with $\bar{n}_c$ replaced with $\bar{n}_c\super{eff} = \bar{n}_c
e^{\operatorname{sign}(p\sub{cav})f\sub{sat}(x)}$, where
$x$ is the combination discussed above that measures the local `anion-ness'.
The SCCS solvation fits for ions required electron density parameters
for anions about an order of magnitude larger than those for cations
and neutral molecules.\cite{PCM-SCCS-charged}
We impose the following conditions on $f\sub{sat}(x)$:
\begin{itemize}
\item $f\sub{sat}(x)=0$ for $x<0$ (cation-like regions)
	to reproduce the similarity of cation and neutral parameters.
\item For $x>0$ (anions), $f\sub{sat}(x)$ saturates to $D\sub{max}$
	for large $x$ so that the modulation of $\bar{n}_c\super{eff}$
	is limited to a factor of $e^{D\sub{max}}$.
	This provides numerical stability.
	We set $D\sub{max} = 3$ which is just sufficient
	to cover the parameter changes observed in the SCCS fits.
\item $f\sub{sat}(x)$ is continuous and differentiable.
\end{itemize}
In order to satisfy these conditions, we select
\begin{equation}
f\sub{sat}(x) = D\sub{max}
\begin{cases}
0, & x \le 0 \\
\tanh x^2, & x > 0.
\end{cases}
\end{equation}
This parametrization is of course not unique, but it is one of the simplest
choices that captures the observed charge asymmetry and remains numerically stable.
Note that a similar dependence on the solute electric field would be extremely unstable
in a conventional isodensity model that depends on the local electronic density.
Here the nonlocality introduced by the convolutions with $w\sub{lq}(r)$
is critical to the success of the present model.

Finally, the fit parameter $p\sub{cav}$ selects the sensitivity of the cavity
to the solute electric field. Water requires $p\sub{cav}>0$ because
anions in water require a larger $\bar{n}_c\super{eff}$ than cations.
Some solvents, such as acetonitrile, exhibit the opposite asymmetry.
Note that we split the sign and magnitude of $p\sub{cav}$ in the
formulation of (\ref{eqn:shapeCANDLE}), so that the charge asymmetry
correction always applies to anions rather than cations.
We could have alternatively applied the correction to anions
when $p\sub{cav}>0$ and to cations when $p\sub{cav}<0$.
However, this choice leads to an instability for cations when $p\sub{cav}<0$:
a decrease in the electron density near the nuclei increases the electric field,
reduces the cavity size, increases the solvation of the electrons, and favors
a further decrease in electron density. (The similar situation of anions
for $p\sub{cav}>0$ is stable because increases in electron density
are limited by the associated Kohn-Sham kinetic energy cost.)

\subsection{Electric response} \label{sec:ElectricResponse}

\begin{figure}
\includegraphics[width=\columnwidth]{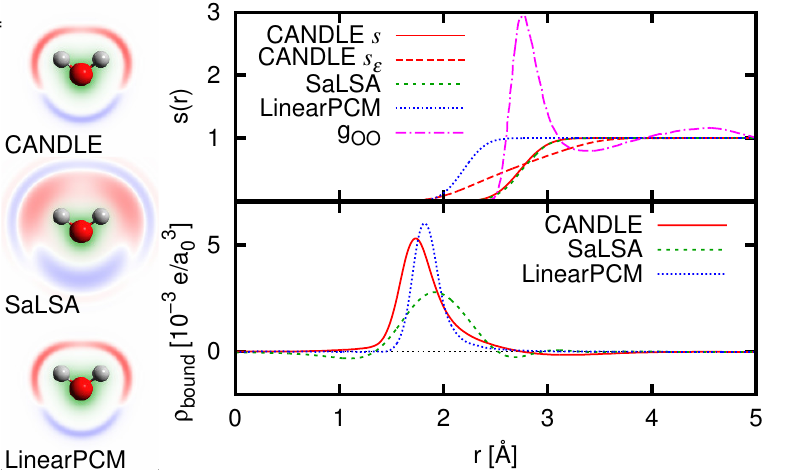}
\caption{Comparison of the cavity shape functions and bound charges
($\rho\sub{bound}(\vec{r})$) for a water molecule in water from the
CANDLE, SaLSA\cite{SaLSA} and local LinearPCM\cite{NonlinearPCM} models.
The experimental oxygen-oxygen radial distribution
function\cite{SoperEPSR} $g_{OO}(r)$ is also shown for comparison.
Note that the SaLSA and CANDLE $s(\vec{R})$ are almost superimposed.
The left panels show the bound charge ($+$ red, $-$ blue) and electron density (green).
\label{fig:nbound}}
\end{figure}

The cavity of conventional density-based solvation models represents the
shape of an effective continuum dielectric that reproduces solvation energies.
In contrast, the cavity of the SaLSA model (and hence the one determined above)
corresponds to the physical distribution of solvent molecule centers because
the model directly captures the nonlocal dielectric response of the solvent molecules.
However, this nonlocal dielectric response requires an expansion in
angular momentum that is computationally intensive and practically applicable
only for solvents involving small, approximately spherical and rigid molecules.

The CANDLE solvation model restores the standard local response approximation
to achieve computational expediency and generality, but this in turn then requires
an empirical description of the dielectric as in other local solvation models.
We use the dielectric shape function
\begin{equation}
s_\epsilon(\vec{r}) = s(\vec{r}) \ast \frac{\delta(r - \eta)}{4\pi\eta^2},
\label{eqn:shapeEpsilon}
\end{equation}
which extends an empirical distance $\eta$ closer to the solute than
the solvent-center cavity $s(\vec{r})$ described by (\ref{eqn:shapeCANDLE}).
The solvent electric response is then approximated by a
continuum dielectric $\epsilon_b$, optionally with Debye screening
$\kappa^2 = \frac{4\pi}{T}\sum_i N_i Z_i^2$ due to finite bulk concentrations
$N_i$ of ions of charge $Z_i$, modulated by the dielectric shape function.
The free energy of interaction of the solute charge density $\rho\sub{el}(\vec{r})$
with the solvent electric response is
\begin{flalign}
U\sub{lq}[\rho\sub{el}] &= \frac{1}{2} \int d\vec{r} \rho\sub{el}(\vec{r})
	\left[ \hat{K}_\epsilon - \hat{K} \right] \rho\sub{el}(\vec{r}), \textrm{ where}\nonumber\\
\hat{K}_\epsilon &\equiv \left[
	\frac{-\nabla\cdot(1+(\epsilon_b-1)s_\epsilon(\vec{r}))\nabla
	+ \kappa^2s_\epsilon(\vec{r})}{4\pi} \right]^{-1}
\label{eqn:ElectricResponse}
\end{flalign}
is the screened Coulomb operator. In practice, $\phi
= \hat{K}_\epsilon \rho\sub{el}$ is calculated iteratively
by solving the modified Poisson (Helmholtz, if $\kappa^2 \ne 0$)
equation, $\hat{K}_\epsilon^{-1}\phi = \rho\sub{el}$,
exactly as in previous solvation models.\cite{NonlinearPCM}

Figure~\ref{fig:nbound} compares the bound charges and cavity
shape functions of the CANDLE solvation model with previous
density-based solvation models, for a water molecule in liquid water.
The SaLSA and CANDLE $s(\vec{r})$ are almost identical and the transition is
at the physical location of the first peak of the radial distribution
function $g_{OO}(r)$. The local LinearPCM requires a cavity that
transitions much closer to the solute, while the CANDLE $s_\epsilon(\vec{r})$
reaches inwards towards the solute with a much wider transition region.
The bound charge in the CANDLE solvation model is qualitatively similar
to the purely local model, except for a longer tail away from the solute
due to the slower variation of the dielectric constant.

\subsection{Dispersion energy} \label{sec:Dispersion}

Finally, for the dispersion energy, we adopt a slightly modified form
of the empirical approximation used in SaLSA,\cite{SaLSA,CavityWDA}
which applies the DFT-D2 pair potential correction\cite{Dispersion-Grimme}
between the discrete solute atoms and a continuous distribution of solvent atoms.
The solvent atom distribution is generated from $s(\vec{r})$ by assuming
an isotropic orientation distribution of rigid molecules.
In order to generalize to non-spherical solvent molecules and eliminate
the dependence on the structure of the solvent molecule, we replace
the atoms in the solvent molecule with a continuous spherical distribution $w\sub{lq}(r)$
of local polarizable oscillators with an empirical effective coefficient $C\sub{6eff}$ each.
The resulting simplified dispersion functional is
\begin{multline}
E\sub{disp}[s] = -\sqrt{C\sub{6eff}}N\sub{bulk} \sum_{i}
	\int d\vec{r} (w\sub{lq}\ast s)(\vec{r}) \\ \times
	\frac{\sqrt{C_{6i}}}{|\vec{R}_i-\vec{r}|^6}
	f\sub{dmp}\left(\frac{|\vec{R}_i-\vec{r}|}{R_{0i}}\right),
\label{eqn:DispersionModel}
\end{multline}
where $N\sub{bulk}$ is the bulk number density of the solvent,
$C_{6i}$ and $R_{0i}$ are the DFT-D2 parameters for solute atom $i$
located at position $\vec{R}_i$, and $f\sub{dmp}$ is the short-range
damping function (see Refs.~\citenum{CavityWDA} and
\citenum{Dispersion-Grimme} for details). The empirical scale factor $s_6$
in the DFT-D2 correction has been absorbed into the empirical $C\sub{6eff}$ coefficient.

\section{Results} \label{sec:Results}

\subsection{Computational details} \label{sec:CompDetails}

We implemented the CANDLE solvation model in the open-source
plane-wave density functional software, JDFTx.\cite{JDFTx}
The local electric response is evaluated iteratively
in the plane-wave basis using exactly the same solver
as previous local solvation models,\cite{PCM-Kendra,NonlinearPCM}
while the nonlocal parts of the functional are shared
with or are minor adaptations of the SaLSA model.\cite{SaLSA,CavityWDA}
The nuclear charge density contributions to $\rho\sub{el}$
are widened to Gaussians so that they are resolvable on
the plane-wave grid (see Ref.~\citenum{NonlinearPCM} for details).
In the calculation of the cavity shape function using (\ref{eqn:shapeCANDLE}),
the valence electron density $n(\vec{r})$ is augmented by $\delta$-functions
that account for all the missing core electrons, to be consistent
with the all-electron convolutions used in the correlation with vdW radii.\cite{SaLSA}

We perform all calculations with the PBE\cite{PBE}
generalized-gradient approximation to the exchange-correlation functional,
and the GBRV ultrasoft pseudopotentials\cite{GBRV} with the
recommended wavefunction and charge-density kinetic energy cutoffs
of 20~$E_h$ and 100~$E_h$ respectively.
At least 15~$a_0$ of vacuum surrounds the solute in each calculation unit cell, 
and truncated coulomb kernels\cite{TruncationMIC,TruncationAnalytic,TruncatedEXX}
are used to eliminate the interaction between periodic images.

\subsection{Parameter fitting} \label{sec:ParameterFitting}

\begin{table}
\begin{tabular}{lcc}
\hline\hline
Parameter & Water & Acetonitrile \\
\hline
\textbf{Fit:} \\
$p\sub{cav}$ [$ea_0/E_h$] & 36.5 & -31.0 \\
$\eta$ [$a_0$] & 1.46 & 3.15 \\
$\sqrt{C\sub{6eff}}$ $\left[\left(\frac{\textrm{J}\cdot\textrm{nm}^6}{\textrm{mol}}\right)^{1/2}\right]$ & 0.770 & 2.21 \\
\hline
\textbf{Physical:} \\
Valence electron count, $Z\sub{val}$ & 8 & 16 \\
vdW radius, $R\sub{vdW}$ [\Angstrom] & 1.385 & 2.12 \\
Dielectric constant, $\epsilon_b$ & 78.4 & 38.8 \\
Bulk density, $N\sub{bulk}$ [$a_0^{-3}$] & \sci{4.938}{-3} & \sci{1.709}{-3} \\
Vapor pressure, $p\sub{vap}$ [kPa] & 3.14 & 11.8 \\
Surface tension, $\sigma\sub{bulk}$ [$E_h/a_0^2$] & \sci{4.62}{-5} & \sci{1.88}{-5} \\
\hline\hline
\end{tabular}
\caption{Fit parameters and physical properties that constrain the CANDLE solvation model.
We obtain vdW radii from Ref.~\citenum{RvdwFluids} and all other physical properties
from Ref.~\citenum{CRC-Handbook} (at standard conditions, $T=298$~K and $p=101.3$~kPa).
\label{tab:Parameters}}
\end{table}

The CANDLE solvation model has three parameters per solvent,
the charge-asymmetry parameter $p\sub{cav}$, the electrostatic radius $\eta$
and the effective dispersion parameter $\sqrt{C\sub{6eff}}$,
that are fit to a dataset of experimental solvation energies
of neutral molecules, cations and anions in that solvent.
Table~\ref{tab:Parameters} lists the optimum fit parameters
for water and acetonitrile that we determine below, along with
values of the physical properties that constrain the solvation model.

We calculate the gas-phase energy for each solute at the optimized vacuum geometry.
We optimize the solution-phase geometry using an initial guess for the
solvation model parameters, and at that optimum geometry, calculate the
solvation energy and its analytical Hellman-Feynman derivatives with respect
to the parameters on a coarse grid in the parameter space of the solvation model.
Using the analytical derivatives, we interpolate the solvation energies
to a finer grid in parameter space and then select the optimum parameters
to minimize the mean absolute error (MAE) of all the solutes.
We re-optimize the solution-phase geometries with these parameters,
and repeat the above parameter sweep process till the optimum parameters converge.
For both solvents considered here, the second sweep yields identical optimum parameters
as the first, and we show the results of that final self-consistent parameter sweep.

\subsection{Water} \label{sec:Water}

\begin{figure}
\includegraphics[width=\columnwidth]{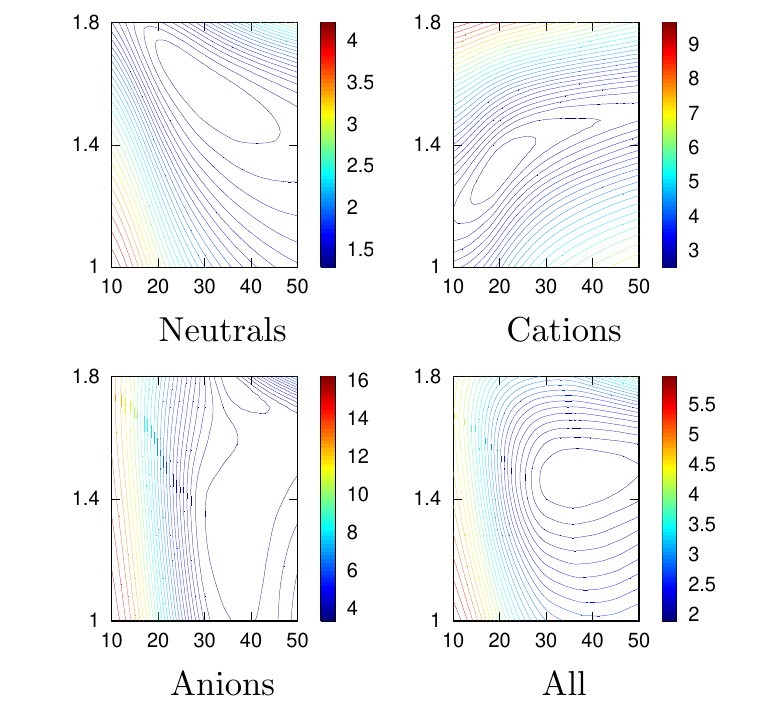}
\caption{Mean absolute error (MAE) of CANDLE solvation energies
of 240 neutral molecules, 51 cations, 55 anions and all of
these solutes in water, as a function of the fit parameters.
In each panel, the $x$-axis is $p\sub{cav}$ in $ea_0/E_h$,
the $y$-axis is $\eta$ in $a_0$ and the contours (color) axis
is MAE in kcal/mol. The $\sqrt{C\sub{6eff}}$ parameter is set
to its optimum value for each combination of the other two parameters.
\label{fig:FitWater}}
\end{figure}

Using the above protocol, we fit the parameters for water to a dataset
of 240 neutral molecules, 51 cations and 55 anions, identical to the
one used in fitting the SCCS models.\cite{PCM-SCCS,PCM-SCCS-charged}
Figure~\ref{fig:FitWater} shows the MAE in the solvation energies
as a function of the solvation model parameters.
Note the extreme sensitivity of the anion solvation energies
to the charge-asymmetry parameter $p\sub{cav}$ ($x$-axis); the MAE
for anions would exceed 15~kcal/mol if $p\sub{cav}$ is set to zero.
The neutral molecules and cations more strongly constrain the
electrostatic radius $\eta$ ($y$-axis). Overall, the net MAE
of all solutes tightly constrains all the parameters.
(The solvation energies depend almost linearly on the
dispersion parameter $\sqrt{C\sub{6eff}}$. 
To simplify the visualization in Figure~\ref{fig:FitWater},
we `integrate out' the $\sqrt{C\sub{6eff}}$ parameter by setting it
to its optimum value for each combination of the other parameters.)

\begin{table}
\begin{tabular}{lcccc}
\hline\hline
\multirow{2}{*}{Model} & \multicolumn{4}{c}{MAE [kcal/mol]} \\
& Neutral & Cations & Anions & All \\
\hline
GAUSSIAN '03       &  --  & 4.00 & 10.2 &  --  \\
GAUSSIAN '09       &  --  & 11.9 & 15.0 &  --  \\
SCCS neutral fit 1 & 1.20 & 2.55 & 17.4 & 3.41 \\
SCCS neutral fit 2 & 1.28 & 2.66 & 16.9 & 3.35 \\
SCCS cation fit    &  --  & 2.26 &  --  &  --  \\
SCCS anion fit     &  --  &  --  & 5.54 &  --  \\
\hline
CANDLE             & 1.27 & 2.62 & 3.46 & 1.81 \\
\hline\hline
\end{tabular}
\caption{Mean absolute errors (MAE) of the CANDLE solvation model
for water compared to various parametrizations of the SCCS model,\cite{PCM-SCCS}
and IEF-PCM\cite{IEF-PCM1,IEF-PCM2} in GAUSSIAN\cite{GAUSSIAN} using identical sets of solutes.
(SCCS and GAUSSIAN results from Ref.~\citenum{PCM-SCCS-charged}.)
\label{tab:WaterAccuracy}}
\end{table}

Table~\ref{tab:WaterAccuracy} compares the accuracy of the
CANDLE solvation model for water with that of the SCCS models
and IEF-PCM\cite{IEF-PCM1,IEF-PCM2} in GAUSSIAN\cite{GAUSSIAN}
on exactly the same set of solutes.
The IEF-PCM model exhibits large errors for cations as well as anions,
while the SCCS model fit to neutral molecules alone, works reasonably
well for cations but systematically undersolvates anions
resulting in a large error of 17~kcal/mol. This error is reduced
to 5.5~kcal/mol by fitting a separate set of parameters for anions alone.
With charge asymmetry built in, the CANDLE solvation model with
a single parameter set exhibits comparable accuracy to
the individual SCCS models fit to each solute type.

\begin{figure}
\includegraphics[width=\columnwidth]{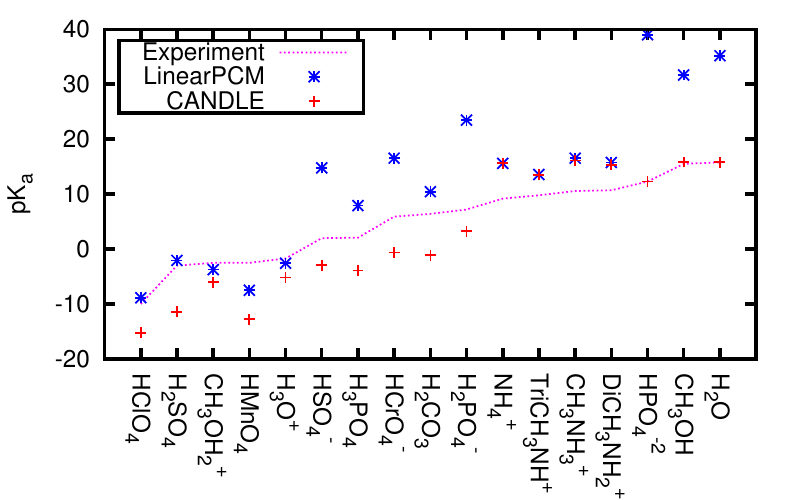}
\caption{Acid dissociation constants (pKa) in water predicted by the
CANDLE and LinearPCM\cite{NonlinearPCM} models compared to experiment.
\label{fig:pKaWater}}
\end{figure}

As an independent test of accuracy, Figure~\ref{fig:pKaWater}
compares predicted acid dissociation constants of mostly
inorganic acids (not present in the fit set) with experiment.
The CANDLE model marginally increases the error in pKa of cationic acids
compared to the local LinearPCM model, but significantly improves the predictions
for neutral and anionic acids since it solves the anion under-solvation issue.
Note in particular that CANDLE makes reasonable predictions even for the
second and third dissociations of sulfuric and phosphoric acid, which require
solvation of dianions and trianions respectively. For the set considered here,
the MAE is 4.7 pKa units for CANDLE compared to 8.4 pKa units for LinearPCM.\cite{NonlinearPCM}

\subsection{Acetonitrile} \label{sec:Acetonitrile}

\begin{figure}
\includegraphics[width=\columnwidth]{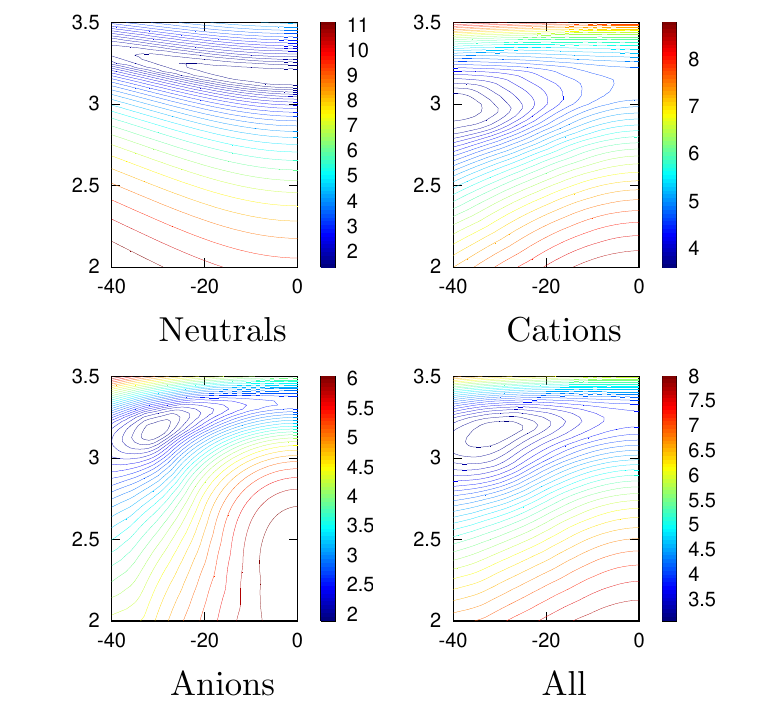}
\caption{Mean absolute error (MAE) of CANDLE solvation energies
of 12 neutral molecules, 30 cations, 39 anions and all of
these solutes in acetonitrile, as a function of the fit parameters.
The axes are exactly analogous to Figure~\ref{fig:FitWater}.
\label{fig:FitMeCN}}
\end{figure}

For acetonitrile, we fit the CANDLE parameters using the above protocol
to the solvation energies of the 12 neutral molecules, 30 cations
and 39 anions in the Minnesota solvation database.\cite{MNsol}
Figure~\ref{fig:FitMeCN} shows the variation of MAE with
parameters for the solvation energies in acetonitrile.
As in the case of water, the combined set of neutral and
charged solutes constrains the fit parameters well.
At the optimum parameters, the MAE is 2.35~kcal/mol for neutral molecules,
4.04~kcal/mol for cations, 1.81~kcal/mol for anions, and 2.97~kcal/mol overall.

In contrast to water, the charge asymmetry parameter is negative for acetonitrile
indicating that cations are solvated more strongly than anions of the same size.
This follows intuitively from the charge distributions of the solvent molecules.
In water, the positively-charged hydrogen sites can get closer to the solute
than the negatively-charged oxygen and hence anions are solvated more strongly.
In acetonitrile, the negatively-charged nitrogen site is more easily solute-accessible
whereas the positively-charged carbon site is blocked by the methyl group,
and therefore cations are solvated more strongly.

\subsection{Solvation of metal surfaces} \label{sec:MetalSurfaces}

Finally, we examine the predictions of the CANDLE solvation model for
a class of relatively clean electrochemical systems: single crystalline
noble metal electrodes in an aqueous non-adsorbing electrolyte.
The surface charge on these electrodes depends on the electrochemical potential,
and the surface becomes neutral at the potential of zero charge (PZC).
Experimentally, these potentials are referenced against the standard
hydrogen electrode (SHE). The absolute level of the SHE is difficult to
determine experimentally and estimates range from 4.4 to 4.9~eV.\cite{PZC}
Correlating the theoretical electron chemical potential of solvated
neutral metal surfaces with the measured PZC, provides a theoretical
estimate of this absolute potential.\cite{PCM-Kendra,NonlinearPCM}
Here, we reexamine this theoretical estimate with the nonlocal
solvation models, CANDLE and SaLSA.

\begin{figure}
\includegraphics[width=0.8\columnwidth]{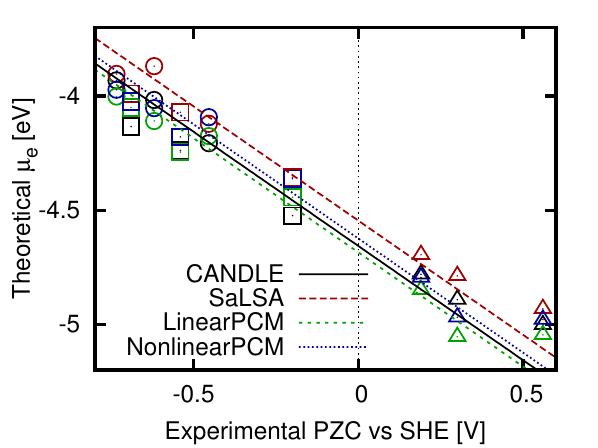}
\caption{Correlation of theoretical electron chemical potentials ($\mu_e$)
with experimental potential of zero charge (relative to standard hydrogen electrode (SHE))
for various solvation models. The results are for single crystalline copper (squares),
silver (triangles) and gold (circles) surfaces, with 111, 100 and 110 orientations
from left to right. (LinearPCM and NonlinearPCM data from Ref.~\citenum{NonlinearPCM}.)
\label{fig:PZC}}
\end{figure}

\begin{table}
\begin{tabular}{cccc}
\hline\hline
Model & $\mu\sub{SHE}$ [eV] & RMS error [eV] \\
\hline
CANDLE       & -4.66 & 0.11 \\
SaLSA        & -4.55 & 0.09 \\
LinearPCM    & -4.68 & 0.09 \\
NonlinearPCM & -4.62 & 0.09 \\
\hline\hline
\end{tabular}
\caption{Offset and RMS deviation between theoretical electron chemical potentials
and experimental potentials of zero charge for various solvation models.
(LinearPCM and NonlinearPCM data from Ref.~\citenum{NonlinearPCM}.)
\label{tab:PZC}}
\end{table}

Figure~\ref{fig:PZC} plots the calculated electron chemical potential of
neutral metal surfaces using various solvation models against the experimental PZC,
and table~\ref{tab:PZC} summarizes the absolute offset and error in the correlation so obtained.
The absolute offsets predicted using various solvation models agree to
within 0.1~eV and are well within the expected experimental range.
The CANDLE model exhibits a marginally higher scatter, but overall agrees well
with the linear and nonlinear local models studied in Ref.~\citenum{NonlinearPCM}.
The nonlocality of the SaLSA and CANDLE models, therefore, does not significantly
alter the predictions of the local solvation models for the absolute SHE potential.

\begin{figure}
\includegraphics[width=\columnwidth]{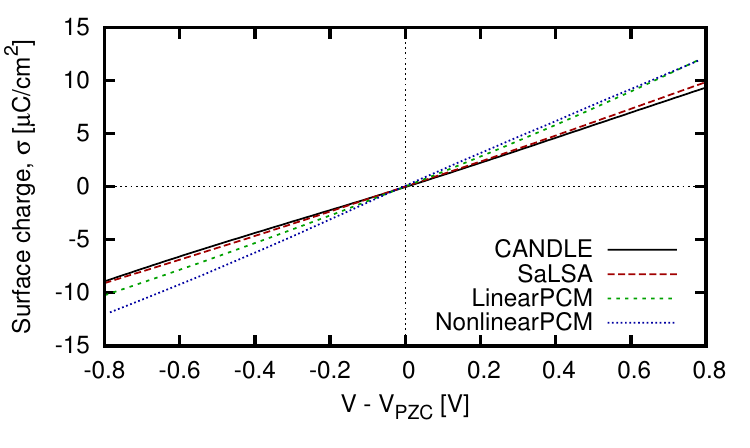}
\caption{Variation of surface charge on the 111 surface of platinum
with electrode potential for various solvation models.
(LinearPCM and NonlinearPCM data from Ref.~\citenum{NonlinearPCM}.)
\label{fig:Charging}}
\end{figure}

The charge of metal electrodes as a function of the electrode potential
is sensitive to the structure of the electrochemical double layer and
varies nonlinearly, but continuum solvation models predict an almost
linear variation (almost constant double layer capacitance).\cite{NonlinearPCM}
Figure~\ref{fig:Charging} shows that the nonlocal solvation models,
CANDLE and SaLSA, also predict a linear charging curve for the Pt 111 surface.
The value of the double-layer capacitance is 12~$\mu$F/cm$^2$
for these nonlocal models, slightly lower than 14 and 15~$\mu$F/cm$^2$
for the linear and nonlinear solvation models\cite{NonlinearPCM}
and an experimental estimate\cite{DoubleLayerCapacitance} of $\sim$20~$\mu$F/cm$^2$.
Details of ion adsorption and the nonlinear capacitance of the electrochemical
interface are therefore not described by continuum solvation models
and require an explicit treatment of the electrochemical double layer.

\section{Conclusions}

This work constructs an electron-density-based solvation model, the CANDLE model,
that explicitly accounts for the asymmetry in solvation of cations and anions.
This model incorporates the charge asymmetry by adjusting the effective
electron density threshold parameter (and hence the cavity size)
depending on the local charge environment of the solute, which in turn
is measured using the direction of the solute electric field on the cavity surface.
The CANDLE model exploits the nonlocal cavity determination and approximations
to the cavity formation and dispersion energies of the fully nonlocal SaLSA model,\cite{SaLSA}
but replaces the nonlocal electric response with an effective local response,
thereby combining the computational efficiency of standard local-response solvation
models with the stability and accuracy of the nonlocal model.

With just three parameters per solvent, the CANDLE model predicts solvation energies
of neutral molecules, cations and anions in water and acetonitrile
with higher accuracy than previous density-based solvation models.
Since a single set of parameters works for differently charged solutes,
the CANDLE model is particularly important for systems that expose strongly-charged
positive as well as negative centers to solution, such as ionic surfaces.
A comparative study of solvation models for solid-liquid interfaces
would be particularly desirable, but difficult due to the dearth of directly
calculable experimental properties (analogous to solvation energies for finite systems).
Constraining the parameters of this model requires experimental solvation energies
for neutral as well as charged solutes, but extensive data is available only
for a small number of solvents. The trends in the CANDLE parameters for other
solvents for which ion solvation data is available will be useful in estimating
the parameters and accuracy of the CANDLE model for solvents without such data.

\section*{Acknowledgements}
We thank Yan-Choi Lam, Dr. Robert Nielsen and Dr. Yuan Ping for suggesting benchmark
systems, for help locating experimental data and for useful discussions.

This material is based upon work performed by the Joint Center for Artificial Photosynthesis,
a DOE Energy Innovation Hub, supported through the Office of Science
of the U.S. Department of Energy under Award Number DE-SC0004993.

%

\end{document}